# Connectivity for AI enabled cities – A field survey based study of emerging economies

*Dibakar Das[1], Jyotsna Bapat[1], Angeliki Katsenou[2] and Sushmita Shrestha[3]*

[1]IIIT Bangalore, India

[2]University of Bristol, United Kingdom

[3] Utopia, Kathmandu, Nepal

**Abstract**

The impact of Artificial Intelligence (AI) is transforming various aspects of urban life, including, governance, policy and planning, healthcare, sustainability, economics, entrepreneurship, etc. Although AI immense potential for positively impacting urban living, its success depends on overcoming significant challenges, particularly in telecommunications infrastructure. Smart city applications, such as, federated learning, Internet of Things (IoT), and online financial services, require reliable Quality of Service (QoS) from telecommunications networks to ensure effective information transfer. However, with over three billion people underserved or lacking access to internet, many of these AI-driven applications are at risk of either remaining underutilized or failing altogether. Furthermore, many IoT and video-based applications in densely populated urban areas require high-quality connectivity. This paper explores these issues, focusing on the challenges that need to be mitigated to make AI succeed in emerging countries, where more than 80% of the world population resides and urban migration grows. In this context, an overview of a case study conducted in Kathmandu, Nepal, highlights citizens' aspirations for affordable, high-quality internet-based services. The findings underscore the pressing need for advanced telecommunication networks to meet diverse user requirements while addressing investment and infrastructure gaps. This discussion provides insights into bridging the digital divide and enabling AI's transformative potential in urban areas.

1. Introduction

Artificial Intelligence (AI) is profoundly reshaping every aspect of human life. AI is transforming humanity by enhancing productivity, personalizing experiences and revolutionizing various industries. In healthcare, AI algorithms analyze medical data to improve screening, diagnosis and treatment plans of patients [1]. In all workplace, AI can automate repetitive tasks enabling employees to focus on more complex high-valued activities which can lead to increased job satisfaction and efficiency [2]. Moreover, AI-powered systems are reshaping daily life through personalized recommendations in entertainment and shopping thereby influencing consumer behavior [3]. AI and robotics together have the potential to automate our various day to day monotonous activities.

AI is increasingly recognized as a powerful tool for improving urban life. By leveraging data driven insights and automation, AI can enhance various aspects of urban living including governance, transportation, public safety, healthcare, energy management and environmental sustainability. By leveraging vast amounts of data generated in urban environments, AI driven solutions enable policymakers to optimize resource allocation, streamline services and increase citizen engagement.

AI has the potential to revolutionize urban transportation systems, making them more efficient and user friendly. By analyzing real-time traffic data, AI algorithms can optimize traffic signal timings, manage congestion, and improve overall traffic flow, leading to reduced travel times and lowers emissions providing a greener solution [4][5]. AI can enhance public transportation by predicting demand and adjusting schedules accordingly ensuring that services meet the needs of commuters [6].

AI technologies can enhance public safety through advanced monitoring and predictive analytics. AI can help in ensuring safety of citizens. AI driven surveillance cameras can detect unusual behavior alerting law enforcement agencies of potential threats and enhance overall security [7]. Predictive policing uses AI algorithms to analyze crime data and identify hotspots allowing police to allocate their resources effectively.

Health care is another sector where AI is playing a very important role, enhancing diagnostics, treatment, and patient care. In medical imaging, AI-powered tools analyze X-rays, MRIs, and CT scans with remarkable accuracy, enabling early detection of conditions like cancer and fractures. Predictive analytics use patient data to forecast disease progression and suggest personalized treatment plans, improving outcomes. Additionally, AI aids in drug discovery by analyzing vast datasets to identify potential compounds, accelerating the development of new treatments. AI enabled platforms facilitate remote consultations, e.g., telemedicine, remote surgery, improving access to healthcare services particularly in underserved areas. All in all, AI can analyze any type of medical data, from imaging to proteomic structures, to predict disease outbreaks and patient needs aiding healthcare providers to prepare and respond effectively [8][9].

AI can provide solutions to optimize energy utilization in urban areas. AI can enhance the efficiency of energy distribution networks, for example, in a smart grid, by predicting demand and managing resources in real time reducing wastage and costs [10]. Intelligent monitoring systems can be designed to monitor the energy consumption in buildings promoting sustainability and reducing utility costs [11]. Data from sensors can be analyzed to provide real-time air quality assessments helping cities to respond to pollution and improve public health [12]. AI can optimize waste collection routes and recycling processes improving efficiency and reducing negative environmental impact [13].

Thus, AI has the potential to comprehensively improve urban life in multiple aspects. However, successful implementation and adoption requires investments in technology, infrastructure and training to ensure that cities can effectively harness the capabilities of AI. As urban areas continue to grow and evolve, AI will play a key part in shaping sustainable and livable environments for urban residents.

Reliable connectivity forms the foundation of urban AI applications. Connectivity means the ability of various devices, systems and persons to connect/exchange information with each other. Much of the smartness of the smart entities is a result of their ability to connect, share and make decisions based on the collective knowledge. Various facets of the smart city paradigm depend heavily on affordable connectivity for all its citizens. It should be noted that over three billion people around the world still do not have access to internet or are underserved [14]. An initiative named "Connecting the Unconnected" has been initiated to bridge this internet access gap [15]. Positive impact of connectivity is evident from two standout instances in India; firstly, the advent of mobile phones with access to internet has enabled information flow among its citizen with increased penetration in underserved and rural areas. Secondly, the unified payment interface (UPI) has revolutionized monetary transactions especially for small business owners. However, in several situations some of these applications fail due to high density of users and inadequate telecommunication networks infrastructure. Many other IoT and video based applications for smart cities require high quality connectivity especially in densely populated areas. With the proliferation of artificial intelligence (AI) for smart city applications, specific Quality of Service (QoS) is expected from the telecommunications networks. From technology perspective, most applications of AI for cities apply distributed algorithms [16]. Thus, they require efficient telecommunication networks for transferring data across different entities in urban domain. Modern 5G and beyond networks are tailored to manage distributed AI applications (e.g. federated learning, edge intelligence etc.) by bringing storage and processing close to the user devices [63].

Developed economies normally drive the deployment of advanced technologies because of its access to resources. Hence, deploying AI applications in urban areas at scale is achievable in such areas. However, emerging economies struggle to meet the aspirations of the users to have high-quality low-cost internet services due to high investment needed to improve the network infrastructure for AI applications. It is evident from the survey which was conducted in Kathmandu, capital of Nepal, which suffers from inadequate network infrastructure to provide broadband internet to the users. Three areas, namely, Asan (a commercial hub), Bansighat (a low income region) and Kusunti (a high income region) were surveyed and the citizens were asked about their experiences of internet services. The common theme from the survey is that the people have aspirations for high speed and reliable internet services for their progress and wellbeing. However, affordability remains a key factor for technology adaptation. Also, inadequate telecommunications infrastructure makes the cause even more difficult.

This paper is organized as follows. Sections 2-8 provides a broad scope of how AI can impact urban life. Sections 9 and 10 discuss the need of high quality telecommunications networks for broadband internet to support AI applications in cities. Section 11 discuss the challenges of leveraging AI for urban areas in emerging countries. A case study of Kathmandu on the current state of internet services is presented in Section 12. Section 13 discusses the challenges of what it will take to make AI for cities to be a reality in emerging economies. Conclusions are draw in section 14.

## 2. AI for Urban Governance

As urban populations continue to increase, cities are facing unprecedented challenges in terms of governance, sustainability and delivery of public services. Artificial Intelligence (AI) has emerged as a powerful tool to address these complexities offering innovative solutions that enhance decision making processes and improve urban living conditions. By leveraging vast amounts of data generated within urban environments, AI driven solutions enable policymakers to optimize resource allocation, streamline services and foster citizen engagement.

Recent advancements in machine learning and data analytics have paved the way for applications that address key urban issues. For example, AI can enhance traffic management through real-time data analysis, reducing congestion and improving air quality [5]. Similarly, AI-driven predictive analytics can assist in urban planning by forecasting housing demand and identifying areas vulnerable to climate change [17]. Furthermore, AI technologies can promote inclusivity by facilitating citizen participation in governance through smart platforms gathering community inputs [18].

However, the integration of AI into urban governance is not without challenges. Ethical considerations, such as, data privacy, algorithmic bias and the transparency of decision-making processes are critical to ensuring that AI solutions serve all citizens equitably [19]. As cities increasingly adopt these technologies, it is essential to navigate these challenges to build more resilient, sustainable and inclusive urban environments.

## 3. AI for Urban Policy and Planning

The integration of Artificial Intelligence (AI) in urban policy and planning is reshaping how cities are managed and developed. By harnessing data driven insights, predictive analytics and machine learning algorithms urban planners can make more informed decisions, enhance service delivery and engage citizens in meaningful ways. AI enables urban planners to analyze large amount of data starting from traffic patterns to socio-economic indicators facilitating data driven decision making. This approach helps in identifying trends and making evidence based policies [20]. Predictive modeling with machine learning can help anticipate urban challenges, such as, traffic congestion, housing shortages and environmental impacts [21]. AI contributes to the development of smart cities by optimizing infrastructure management. This includes smart traffic lights, waste management systems and energy-efficient buildings that respond dynamically to real-time conditions [22]. AI tools can enhance public participation by providing platforms for feedback and engagement. Data analytics can help identify citizen needs enabling more inclusive decision making processes. AI applications are vital for sustainable urban development. By analyzing environmental data, cities can develop strategies to mitigate climate change impacts and enhance resilience to those [23].

Despite the potential benefits, several challenges accompany the deployment of AI in urban policy and planning. The collection and use of vast amounts of data raise concerns about privacy and security. Ensuring robust data protection measures is essential to maintain public trust. AI systems can unintentionally propagate biases present in historical data leading to inequitable outcomes in urban policy. Addressing these biases is critical for fair governance [24]. Successful AI integration requires collaboration among urban planners, data scientists, ethicists and citizen stakeholders to ensure that solutions are effective and equitable. The rapid pace of AI development necessitates adaptive regulatory frameworks that can address the ethical and social implications of AI in urban contexts.

AI-driven urban policy and planning hold great promise for creating smarter and more sustainable cities. By leveraging data and advanced analytics, urban planners can enhance decision making processes and engage citizens. However, addressing the ethical, social and technical challenges is crucial to ensure that these innovations lead to equitable and inclusive urban environments.

## 4. AI in Urban Healthcare

The integration of Artificial Intelligence (AI) in urban healthcare is revolutionizing the way health services are being delivered, managed and optimized in urban areas. With the increasing complexity of urban health challenges, such as, rising chronic diseases, resource constraints and the need for timely responses to public health emergencies, AI technologies offer innovative solutions that enhance patient care, improve operational efficiencies and promote healthier communities.

AI algorithms can analyze large datasets to identify trends and predict outbreak of diseases. By utilizing historical health data, demographic information and environmental factors these systems can forecast the spread of infectious diseases enabling urban health departments to implement proactive measures [25]. AI driven telemedicine platforms provide remote consultations making healthcare more accessible to underserved urban population at a lower cost. AI can analyze individual patient data, such as, genetic information, lifestyle factors and medical history to develop tailored treatment and health insurance plans. This personalized approach improves treatment efficacy and patient outcomes especially for chronic conditions prevalent in urban settings [26].

Virtual health assistants powered by AI can triage patients, provide medical advice and follow up on treatment plans thereby improving patient engagement and adherence to health care services [9]. AI applications can optimize scheduling, resource allocation and patient flow within urban healthcare centers and hospitals. Machine learning algorithms can predict patient volumes allowing hospitals to allocate staff and resources more effectively [27]. AI can enhance surveillance systems by processing data from various sources including social media, emergency room visits and health reports. This real-time data analysis helps public health officials identify emerging health threats and also respond swiftly to such emergencies [28].

Though, AI holds great promise for urban healthcare, several challenges have to be addressed. Ensuring the confidentiality of patient data is of paramount importance. Robust security measures must be implemented to protect sensitive health information. AI systems can have biases present in the data they are trained on which may lead to disparities in healthcare delivery. Ongoing efforts to ensure fairness and equity in AI applications are crucial [29].

AI applications in urban healthcare represent a significant scope for improving health outcomes and operational efficiency. By leveraging predictive analytics, telemedicine, personalized treatment and enhanced public health surveillance, cities can address the unique healthcare challenges. However, it is essential to deal with the ethical and practical challenges associated with these technologies to ensure that they serve all communities equitably.

5. AI for Sustainable Environment

The urgent need to address climate change has enabled the application of Artificial Intelligence (AI) into renewable energy solutions, emissions reduction strategies and urban heat island management. By optimizing energy generation, improving efficiency and predicting demand, AI technologies are transforming how cities approach sustainability and efficient energy consumption.

AI can significantly enhance the efficiency and effectiveness of renewable energy systems particularly in optimizing energy generation from sources like solar and wind. AI algorithms can analyze data from sensors installed on renewable energy equipment to predict failures and schedule maintenance thus minimizing downtime and maximizing green energy output [30]. Machine learning models can forecast energy generation based on weather conditions which aids grid management and improves integration of renewable sources [31].

AI technologies are pivotal in monitoring and reducing greenhouse gas emissions from urban areas. They enable cities to track emissions in real-time and identify polluting sources which is crucial for effective climate action. AI can process satellite imagery and data from sensors to monitor air quality and emissions in real time helping policymakers take effective action [32]. AI can optimize traffic flow and public transportation systems reducing emissions from vehicles and improving overall urban mobility [4][5][6].

Urban heat islands (UHIs) are areas in cities that experience significantly higher temperatures than their rural surroundings largely due to human activities and infrastructure. AI can help mitigate these effects. AI techniques can analyze satellite images and ground data to create detailed heat maps of urban areas [33]. By integrating AI into urban planning, cities can design green spaces and implement cooling strategies that mitigate UHI effects. AI models can simulate various scenarios to determine the most effective interventions to reduce the impact of UHI [34].

The deployment of AI systems for urban applications requires significant amount of energy. Ensuring that this demand is met sustainably is crucial. Research is underway to develop energy-efficient AI algorithms that reduce the overall energy demand of AI systems particularly in data centers [35][36]. AI itself can facilitate the integration of AI systems with renewable energy sources ensuring that the additional energy demand from AI is sourced sustainably [37].

AI applications in renewable energy solutions, emissions reduction and urban heat management are crucial for building sustainable urban environments. By optimizing energy usage, enhancing the efficiency of renewable resources and mitigating the effects of urban heat islands, AI technologies can significantly contribute to climate action and sustainable urban development. Addressing the energy demands of AI systems within these contexts ensures that the benefits of AI are realized without compromising environmental goals.

## 6. Ethical AI for Urban Environments

As Artificial Intelligence (AI) technologies become increasingly integrated into urban environments, the need for robust legal frameworks to govern their deployment and use has become crucial. These frameworks aim to ensure that AI systems are developed and operated in ways that are ethical, transparent and accountable, particularly in the context of unique challenges and complexities posed by urban settings.

The rapid adoption of AI in urban settings spanning areas, such as, transportation, public safety, healthcare and urban planning, raises significant legal and ethical issues. Key concerns include data privacy, algorithmic bias, accountability for decision-making and the impact of automation on employment. Effective legal frameworks can mitigate these risks and promote the responsible use of AI technologies. [38].

Several jurisdictions have begun to establish legal frameworks specifically addressing AI technologies. The EU is at the forefront of AI regulation with its proposed AI Act, the European Union AI Act, which categorizes AI applications based on risk levels (high-risk, limited-risk and minimal-risk). High-risk applications, such as, those used in critical infrastructure or biometric identification have to comply stricter requirements regarding transparency, accountability, and safety [39]. Though, not exclusively focused on AI, the General Data Protection Regulation (GDPR) sets out stringent data protection and privacy requirements that apply to AI systems processing personal data. This includes principles of transparency, data minimization and individuals' rights to access and delete their data [40].

Various states and municipalities in United States have initiated their own regulations. For instance, California's Consumer Privacy Act (CCPA) emphasizes consumer rights over personal data which has implications for AI systems utilizing these data [41]. The United Nations (UN) has recognized the need for ethical AI governance particularly in urban settings. The "AI for Good" initiative encourages the development of global standards for AI to promote sustainable urban development [42].

For developing legal frameworks for AI in urban environments, several key considerations should be taken into account. Frameworks should include ethical guidelines that address algorithmic bias, transparency in decision-making and accountability for AI-driven actions [43]. Engaging with stakeholders, including citizens, technologists and ethicists, is crucial to ensure that regulations are grounded in societal values and address public concerns. Given the global nature of AI technologies, international collaboration is essential to harmonize regulations and share best practices [44].

Developing ethical frameworks for regulating AI in urban environments is critical to harnessing the potential of these technologies to mitigate risks. By focusing on ethics, accountability and public engagement, and drawing upon international best practices, policymakers can create robust regulations that promote responsible use of AI in cities. As urban areas continue to evolve, these frameworks will play a vital role in ensuring that AI contributes to sustainable and equitable urban development.

## 7. AI Impact on Urban Labor Markets

As cities evolve into increasingly complex ecosystems, the integration of robotics and artificial intelligence (AI) is reshaping urban environments and labor markets. Urban robotics, encompassing a range of automated systems from delivery drones to autonomous vehicles plays a pivotal role in enhancing urban infrastructure and services. However, the proliferation of AI technologies also presents significant implications for employment patterns and workforce skills development.

Urban robotics refers to the application of robotic systems within city settings to improve efficiency, safety and service delivery. Robotics technology facilitates the delivery of goods through autonomous vehicles and drones reducing delivery times and costs. This innovation has been particularly beneficial during crises, like, the COVID-19 pandemic, where contactless delivery became essential [45]. Robots are increasingly used in public spaces for various services including maintenance, security and information dissemination. This not only enhances operational efficiency but also improves the citizen experience [46].

The integration of AI technologies into urban environments has significant implications for labor markets. Though, AI has the potential to automate tasks traditionally performed by humans leading to job displacement in sectors like manufacturing and logistics, it also has the potential to create new job opportunities in technology development, maintenance and oversight [2]. The rise of the gig economy facilitated by AI platforms is reshaping traditional employment patterns. Workers increasingly engage in short-term and flexible jobs rather than permanent positions leading to shifts in job security and benefits [47].

As AI and robotics continue to transform urban labor markets, the demand for new skills has become significant. There is an increasing need for technical expertise in AI and robotics including programming, data analysis and system maintenance. Educational institutions are adapting curricula to meet these demands [48]. As automation takes over routine tasks, soft skills, such as, creativity, problem-solving and interpersonal communication are becoming increasingly valuable. Employers are seeking workers who can complement AI technologies rather than compete with them [50].

The integration of urban robotics and AI into city environments presents both challenges and opportunities for urban labor markets. Though automation may displace certain jobs, it also creates new roles and demands for skills that are essential in an increasingly automated world. Policymakers, educational institutions and businesses must collaborate to ensure that workforce development initiatives address these changing needs equipping individuals with the skills necessary to thrive in the evolving labor market. As cities embrace these technologies, a proactive approach to workforce transition will be critical in fostering inclusive and sustainable urban development.

### 8. AI for Economics in Urban Areas

Artificial Intelligence (AI) is increasingly recognized as a catalyst for economic growth especially in urban areas where innovation and entrepreneurship thrive. As cities grapple with challenges, such as, population density, resource management and infrastructure demands, AI offers transformative solutions that enhance efficiency, drive new business models and promote sustainable development.

AI technologies enable businesses to optimize operations and improve productivity. Automation of routine tasks, data analysis and predictive analytics empower companies to make informed decisions quickly and efficiently. This increased efficiency not only lowers operational costs but also allows businesses to allocate resources more strategically. AI serves as a powerful tool for entrepreneurs by lowering entry barriers and enabling the development of innovative business models. Startups can leverage AI technologies to create solutions that address urban challenges, such as, transportation, healthcare and environmental sustainability [49].

AI fosters an environment conducive to innovation by providing tools that enable research and development across various sectors. Urban areas which are often home to universities and research institutions, benefit from AI by facilitating collaboration between academia, industry and government. AI-driven research projects in urban areas focus on smart city technologies, climate resilience and public health innovations fostering a culture of innovation that attracts talent and investment [51].

Although concerns about job displacement due to AI are prevalent, the technology also creates new job opportunities particularly in sectors that require advanced technical skills. Urban areas can capitalize on this by developing training programs that equip the workforce with relevant AI skills. Programs focused on AI and data science are essential to prepare workers for the evolving job market. Collaborations between educational institutions and technology companies help bridge the skills gap [2].

AI technologies play a crucial role in addressing urban challenges related to sustainability. By optimizing resource usage, reducing emissions and enhancing public services, AI contributes to the economic development of smart cities. AI applications in energy management, waste reduction, and transportation planning are vital for creating more sustainable urban environments. Smart grids and AI-based traffic management systems exemplify how technology can enhance sustainability [5][10][11].

AI is a powerful driver of economic growth, entrepreneurship and innovation in urban areas. By enhancing productivity, stimulating new business models, fostering innovation, creating jobs and promoting sustainability, AI technologies are reshaping the urban economic scenario. For cities to fully leverage these benefits, it is crucial to invest in education and training, facilitate collaboration between stakeholders and create an environment that supports innovation for economic development. As urban areas continue to evolve, embracing AI will be key to building resilient and prosperous communities.

## 9. Networking for AI (in urban areas)

As artificial intelligence (AI) continues to advance, the integration of networking technologies plays a crucial role in supporting urban AI applications. Networking for AI encompasses the infrastructure, protocols and communication methods required to facilitate data transfer, model training and deployment across distributed systems [52], typical of smart city applications.

The performance of AI systems heavily depends on the availability of data and the computational resources required for training complex models. Rapid access to large datasets is essential for training AI models. Distributed training techniques leverage multiple nodes to enhance processing speed and reduce latency. AI applications increasingly deploy models at the edge necessitating robust networking to manage real-time data processing.

Modern AI architectures often involve a combination of *cloud, fog, mist* and *edge* computing requiring robust networking strategies [53][54][55][56]. Low-latency connections are critical for real-time AI applications, such as, autonomous vehicles and smart cities. Protocols like MQTT and gRPC optimize data transmission, enhancing the performance of AI applications. Networking solutions must accommodate the growing demands of AI workloads especially during peak hours.

The rollout of 5G technology offers higher bandwidth and lower latency significantly benefiting AI applications that require real-time data processing [57]. Network Function Virtualization (NFV) allows the deployment of AI functions on virtualized networks improving flexibility and resource utilization. This decentralized approach to machine learning, e.g., federated learning, leverages distributed devices for training necessitating secure and efficient networking to synchronize model updates [58]. As AI systems often handle sensitive data, ensuring robust security in networking is of paramount importance. Techniques, such as, encryption, secure multi-party computation and blockchain can enhance the privacy of AI data transactions [59].

Networking is a foundational component for the successful deployment and scaling of AI applications for smart cities. As technologies evolve, the interplay between networking and AI will become increasingly significant which will drive innovations across various sectors.

## 10. Cost-Effective Telecommunications Networks for Deploying AI in Urban Areas

The deployment of Artificial Intelligence (AI) systems in urban areas is significantly influenced by the quality and cost-effectiveness of telecommunications networks. High-quality telecommunications infrastructure facilitates the rapid transmission of data which is critical for AI applications ranging from smart city initiatives to real-time analytics.

AI systems rely on vast amounts of data collected from various sources including IoT devices, sensors and user interactions. Effective telecommunications networks ensure seamless connectivity and data transfer which are crucial for the real-time functioning of AI applications. Telecommunications networks with low latency and high bandwidth allow for quick data exchange, enabling AI systems to process information in real time. This is particularly important for applications like traffic management and emergency response [53].

Cost-effective telecommunications infrastructure lowers entry barriers for deploying AI solutions. Municipalities and businesses can invest in AI technologies without the burden of high connectivity costs making innovative solutions more accessible. Collaborations between public entities and private telecommunications providers can lead to the development of affordable network solutions enabling cities to implement AI-driven applications for smart lighting, waste management and public safety systems [60].

AI systems often rely on Internet of Things (IoT) devices to gather data and make informed decisions. A robust telecommunications network supports the extensive connectivity required for IoT deployments which in turn enhances the capabilities of AI systems. Cost-effective networks facilitate the scalability of IoT applications allowing cities to expand their smart city initiatives without significant investment in new infrastructure [61].

The availability of real-time data through efficient telecommunications networks empowers city planners and policymakers to make data-driven decisions. AI systems can analyze this data to optimize urban services and enhance quality of life. By integrating AI with telecommunications infrastructure, cities can deploy analytics platforms that improve services, such as, public transportation, energy management and environmental monitoring [62].

Cost-effective telecommunications networks promote collaboration among various stakeholders including government, businesses and research institutions. Such collaboration is essential for fostering innovation in AI applications. Cities with strong telecommunications infrastructure can become innovation hubs, attracting startups and tech companies focused on developing AI solutions tailored to urban challenges [63].

Improved and cost-effective telecommunications networks are vital for the successful deployment of AI systems in urban areas. By enhancing data transmission, reducing costs, supporting IoT ecosystems, enabling data-driven decision-making and fostering collaborative innovation, robust telecommunications infrastructure lays the groundwork for smart city initiatives and sustainable urban development. As cities continue to embrace AI, investing in telecommunications will be crucial for realizing the full potential of these transformative technologies for urban settings.

## 11. AI for Urban Areas of Emerging Countries

As emerging countries grapple with rapid urbanization, the integration of Artificial Intelligence (AI) into urban planning and management offers transformative opportunities to improve quality of life. However, the successful deployment of AI systems necessitates a robust telecommunications and computing infrastructure. AI-driven solutions can optimize traffic flow, reduce congestion and enhance public transport efficiency even in underdeveloped transport infrastructure. Predictive analytics can improve route planning and scheduling making transit systems more reliable [6]. AI applications in healthcare can improve disease prediction and management through data analysis and machine learning enabling better resource allocation in urban health systems in emerging economies [25][26]. AI can enhance waste collection processes through smart bins that notify collectors when they are full optimizing collection routes and reducing operational costs to compensate for inadequate waste management [13].

Robust telecommunications infrastructure is crucial for the effective deployment of AI systems. Reliable and high-speed internet is essential for data transmission and connectivity among IoT devices. Expanding broadband access in urban areas can facilitate the integration of AI technologies [64]. Leveraging mobile network infrastructure can enhance connectivity particularly in regions where fixed-line networks are less developed. Mobile platforms can support various AI applications in urban management.

Despite the potential benefits, several challenges hinder the effective use of AI in urban areas of emerging countries. Many urban areas lack the necessary telecommunications infrastructure to support advanced AI technologies which can limit data collection and processing capabilities. Concerns about data privacy and security are heightened in emerging countries where regulatory frameworks may be weak or nonexistent. There is often a shortage of skilled professionals capable of developing and managing AI systems which can impede implementation [65].

To overcome these challenges, a multi-faceted approach is required. Governments and international organizations should prioritize investments in telecommunications and computing infrastructure to ensure that urban areas can support AI technologies. Public-private partnerships can be effective in funding and developing necessary infrastructure. Training programs should be implemented to develop local talent in AI and data analytics. Collaborations with universities and tech companies can enhance workforce skills [66]. Establishing adequate regulatory frameworks for data privacy and security is crucial to building trust in AI systems. Governments should collaborate with stakeholders to create policies that protect citizens while promoting innovation [67].

The integration of AI in urban areas of emerging countries holds significant promise for improving quality of life but it is dependent on establishing a solid telecommunications and computing infrastructure. By addressing challenges, such as, infrastructure deficits, data privacy concerns and skill gaps, these countries can harness AI's potential for sustainable urban development. Collaborative efforts among governments, businesses and educational institutions will be vital in driving these initiatives forward ultimately contributing to enhanced living conditions and economic growth.

## 12. Telecommunications Infrastructure in Emerging Economies – Kathmandu Case Study

Having understood the role of telecommunications infrastructure for the success of AI in urban areas generally and more specifically the associated challenges in emerging economies, this paper now presents a study of connectivity issues of Kathamandu, the capital of Nepal and a typical metropolitan of a emerging country.

Urban cities like Kathmandu are experiencing rapid growth, increasing the demand for reliable network infrastructure. As more businesses and transactions are turning digital, the need for accessible, affordable and reliable networks has increased. Many areas in Kathmandu suffer from poor connectivity, patchy network due to frequent outages and lack of access to affordable and reliable services. This issue disproportionately impacts poor and marginalized communities living in densely populated, flood-prone, and geographically challenging areas.

To address this issue, University of Bristol (UoB), United Kingdom (UK) and International Institute of Information Technology Bangalore (IIITB), in partnership with Utopia, Kathmandu and funded by Royal Academy of Engineering, UK, conducted a study to provide fast, reliable networks enabling users to share information and communicate.

The objective of the project was to:

1. Deep dive into the connectivity issue in Kathmandu
2. Design intervention based on the research outcome
3. Socio-technical feasibility study

Three different types of locations were chosen in Kathmandu – a commercial hub with high population density, a low-income settlement and a mid to high income settlement area. Figure 1 shows the age distribution of the survey with majority of the subjects 40 years or less. Education levels of the population sample is depicted in Figure 2 which is dominated by school level education. Gender distribution is shown in Figure 3 which has 42% as males and rest females. Figure 4 shows the occupations distribution of the sample which includes student, homemakers and shop owner.

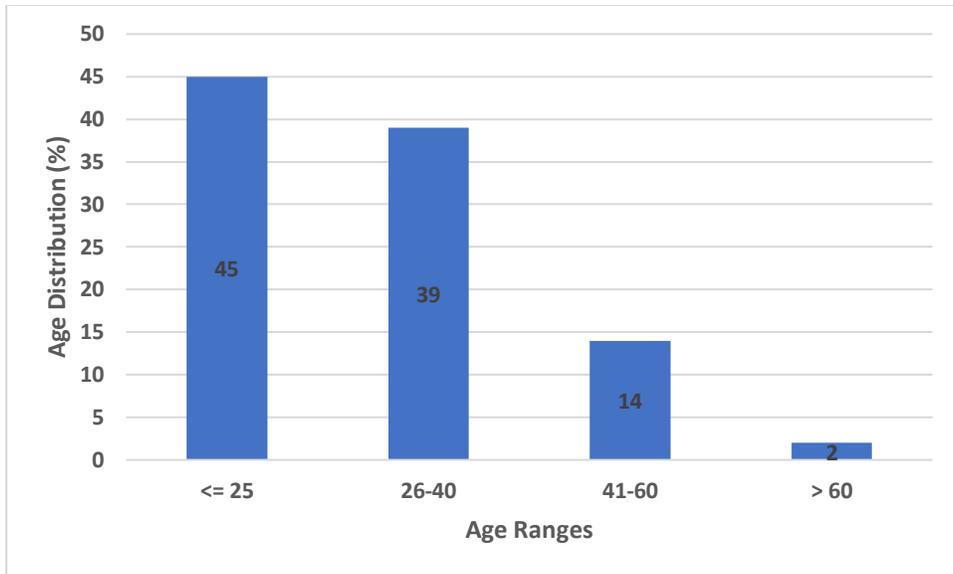

*Figure 1: Age Distribution*

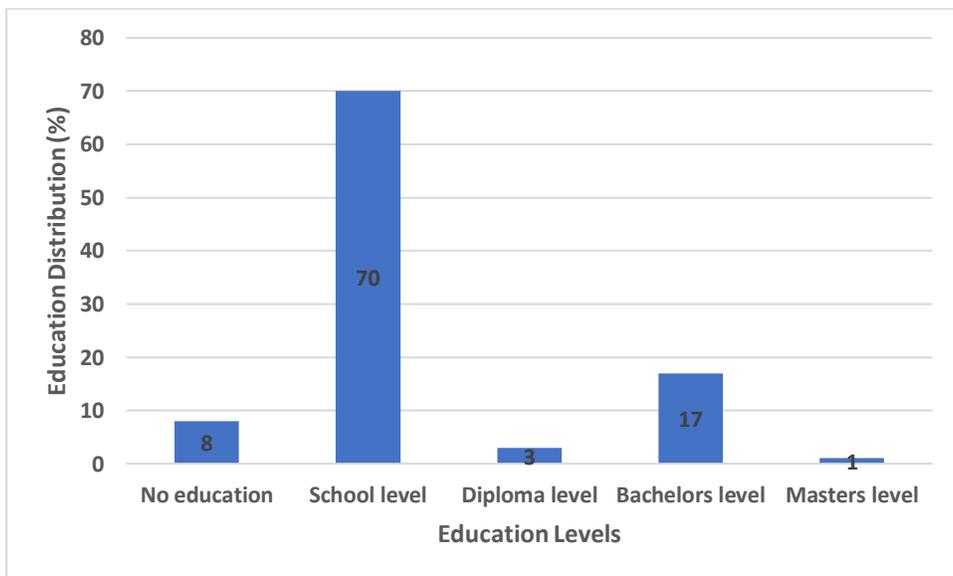

*Figure 2: Education Distribution*

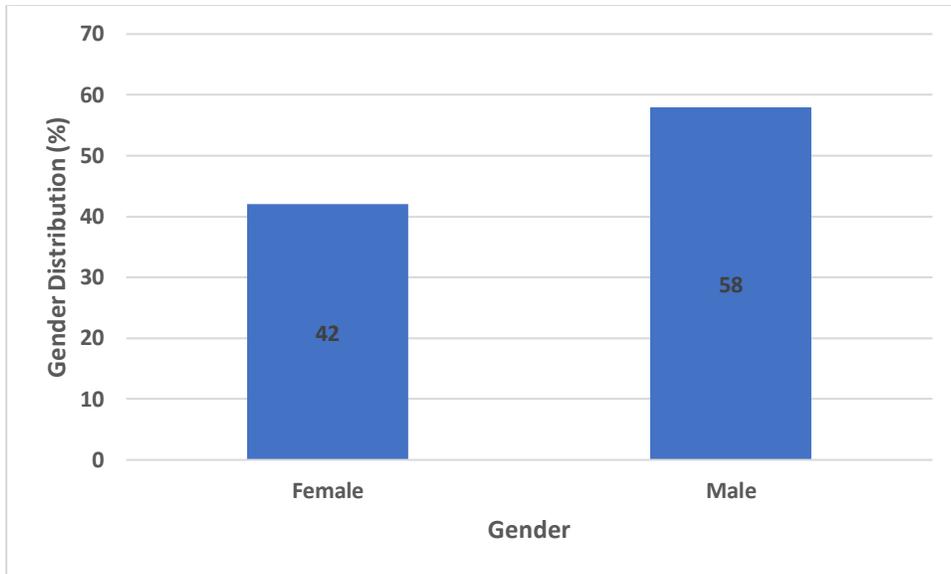

*Figure 3: Gender Distribution*

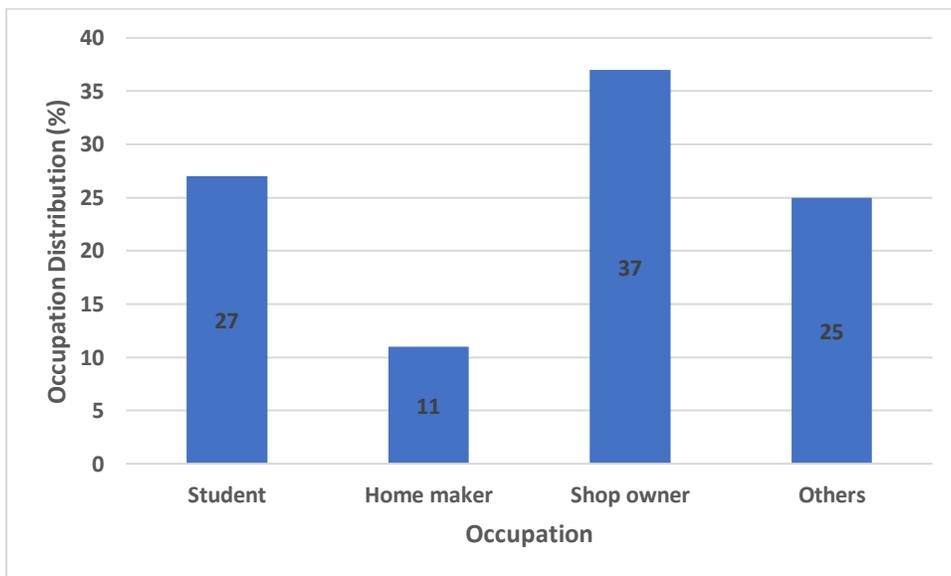

*Figure 4: Occupation Distribution*

### 12.1 Commercial Hub (Asan)

Commercial hub is a busy, bustling street with narrow alleyways lined with commercial establishments run by small business owners. This location was selected for the survey due to its high population and housing density, its compact spatial layout and the rapid transformation of businesses using digital transactions.

Primarily applications used in this area are communication, entertainment and digital banking. Majority of the interviewees were small business owners who use the internet to communicate with clients, suppliers, etc. Interviewees use the internet for entertainment to pass time during the day when they are free. Interviewees use Facebook to watch videos, Facebook messenger to talk to their family members abroad, watch news online and YouTube to watch movies and listen songs. The younger population below 20 years use the internet to play games on their mobile devices. One-third of them use mobile banking applications. Half of them use Facebook and YouTube to stay informed with the news. People between 19-22 years use Instagram to view reels.

One of the major challenges in this area was many users connect to one Wi-Fi router leading to network congestion. They complained about disturbance in connectivity due to heavy traffic in the network especially during evenings. They also complained about slow buffering of mobile banking.

Interviewees also experienced sudden slowdown of internet services. They felt irritated when faced with this problem and tried to turn the Wi-Fi -router on and off, change password or use a cellular data package (which is costly) from their network provider. Sudden slowdown occurred when many people are viewing Instagram reels, playing games, using mobile banking and uploading pictures on social media. A few of the interviewees experienced lack of network connectivity in narrow alleyway surrounded by tall buildings. Also, some of them mentioned about lack of internet connectivity during natural disasters.

On the aspiration front, majority of the people hope for free Wi-Fi, better services and high speed connectivity. They prefer portable and wireless networks so that the city would look clean without wires hanging from the poles.

### 12.2 Urban Low-income Area (Bansighat)

Majority of respondents used the internet for entertainment (e.g., movies, series, games, songs). Next important usage of the internet was personal and professional communication. Respondents use apps like Facebook, Viber, WhatsApp to stay connected with their family, friends and relatives. One user shared that the internet has helped her fight depression as it keeps her connected with her family and friends. Also, people use internet for online payments.

Slow down and buffering are major problems especially during video streaming (during morning, night and weekends) due to higher number of users. This problem leads to decrease in work efficiency due to underperforming internet services (e.g., unable to send emails, do online research and attend online meetings) and also disturbs online classes. Wires of all the ISPs are on the same pole making it difficult to maintain the wirings. Hacking attempts and spam messages are some of the emerging challenges. Phishing websites hack social media and bank accounts.

On average, users are paying NPR 1430/month for the wired internet (via Wi-Fi access point) and they wish to get internet service at cheaper rates. Cellular data service costs vary based on different packages which majority of users find expensive, therefore Wi-Fi based broadband connectivity is preferred over cellular data services. Majority of the users complained about the slow internet. Hence, they wish for lag free fast internet. Also, wireless last mile connectivity to avoid wiring across the areas is also an aspiration. Users wish to see proper security and privacy protection, and bug free and glitch free apps and internet services.

### 12.3 Urban High-income Area (Kusunti)

All users used the internet for entertainment (movies, series, games, songs) during leisure. Users use apps like Facebook, Viber, WhatsApp to stay connected with their family, friends and relatives. People mentioned using internet to decide when and where to purchase products based on discounts. Couple of interviewees mentioned using the internet to learn cooking. Female respondents mentioned using the internet for informal education. Some of the respondents used the internet for research on their respective areas of study. Significant number of interviewees used the internet for mobile banking. Some of the users mentioned that the internet helped their businesses with online marketing and transactions.

Slow internet remains the main problem. Some of the respondents mentioned that the Internet is not reliable and slow down during important works which was very frustrating. Internet is slow during peak usage hours. High-definition video does not play smoothly. High cost of internet is also reported.

Similar to other two areas, cheap internet services through Wi-Fi are preferred as cellular data is expensive. High-speed internet is a necessity to all the respondents. They also look for protection of their online data and transactions.

### 12.4 Synopsis of Survey

Figure 5 and Figure 6 shows the word cloud (nouns, adjectives and verbs) and frequencies of the top 25 words (nouns, adjectives and verbs) respectively. The story these plots try to say is that there are large number of users of internet who value its utility. However, they face many issues with the services, such as, slow video streaming, disruptions due to disasters, network congestion related issues, etc. They prefer Wi-Fi based internet as it is cheaper compared to cellular data. Pricing of internet services remains an issue. They aspire to have high-speed, free and secure internet service supported by the government.

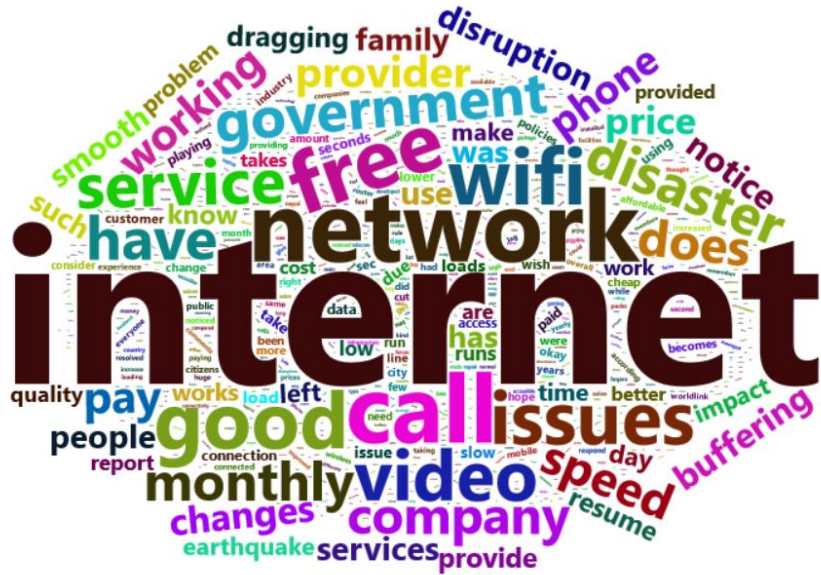

Figure 5: Word cloud of survey

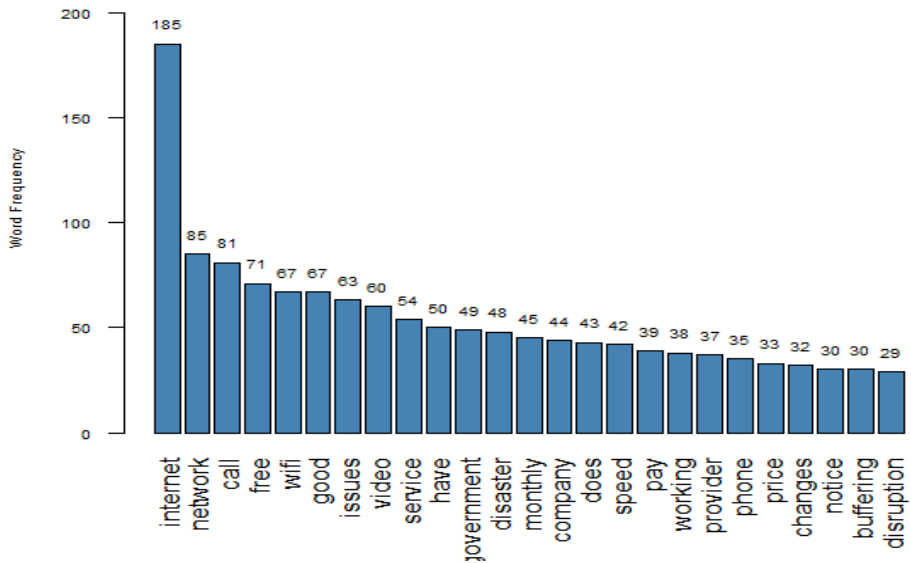

Figure 6: Frequencies of 25 words

12.5 Techno-economic simulation of bandwidth utilization and network congestion

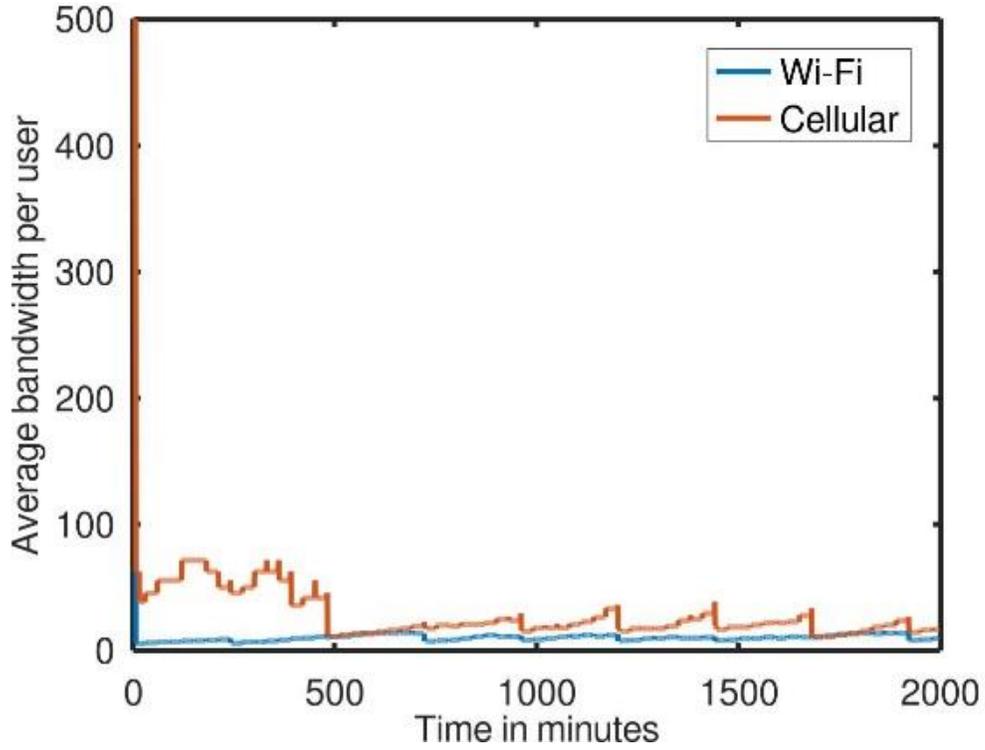

*Figure 7: Wi-Fi network congestion due to low cost*

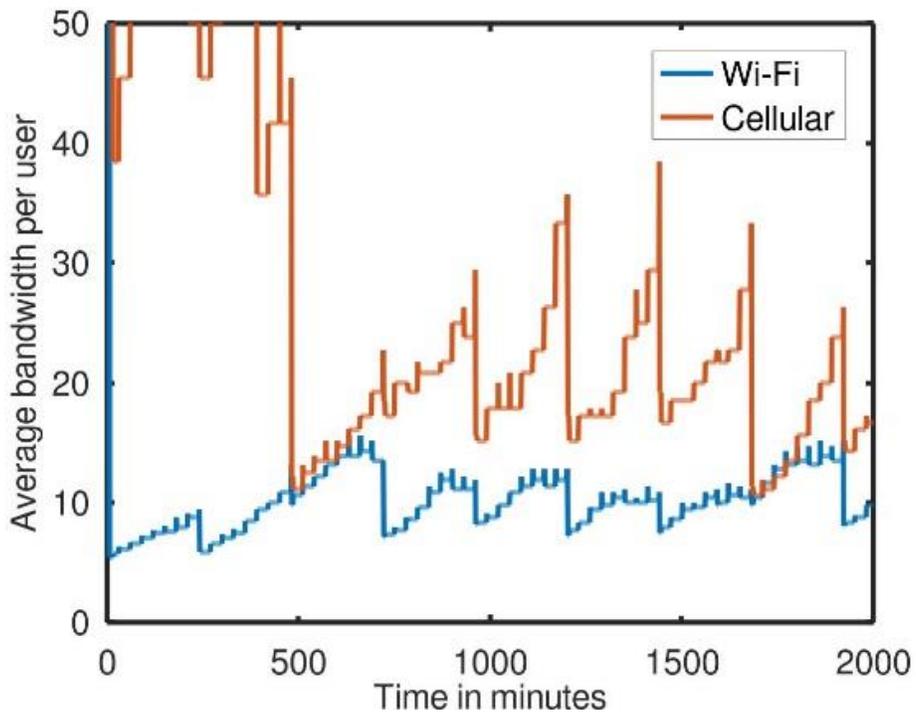

*Figure 8: Wi-Fi network congestion (zoomed version)*

During the survey, it was evident that the price of cellular data, e.g., LTE, is much higher than Wi-Fi network. This meant users preferring Wi-Fi over cellular which led to network congestion. Figure 7 shows a simulation of this congestion of Wi-Fi network as observed. Cellular data was assumed to be priced at ten times that of Wi-Fi. Network congestion is declared when the average bandwidth per user is less than 10 Mbps. The zoomed version of the episode is shown in Figure 8 where it can be observed that at several places the Wi-Fi bandwidth (blue line) is less than 10 Mbps whereas the cellular bandwidth is always better. Situation improves temporarily when users retreat from Wi-Fi due to acute congestion. However, the same cycle repeats because the lower cost drives users again back to Wi-Fi network after some time. The phenomenon was observed in the deployed network as well during the survey. This result shows that proper pricing is internet services is key to avoid network congestion since the latter can have negative impact the essential services provided in urban areas.

13. Discussion

The common theme of the survey is that the people of emerging economies have high aspirations with high speed and reliable internet services for their progress and wellbeing. However, affordability remains a key factor for technology adaptation. Also, inadequate telecommunications infrastructure makes the cause even more difficult. There is a need for huge investments and at the same time return on investments in this sector. As seen from the previous discussion, high quality internet services with modern telecommunication network are fundamental for the success of AI applications in urban areas. Thus, a huge gap must be filled between the current state of affairs in emerging economies to the reality of modern AI enabled urban life. Since, access to internet is no longer a luxury but a basic necessity, building necessary infrastructure [69] through pubic (or public private) funding is probably the way forward. Also, providing free internet for essential services to the users would go a long way to bridge the digital divide. In this context, network slicing [68] and innovative methods of spectrum sharing [70][71][72] can play an important role in ensuring that the required QoS is always met for the essential services to be provided by over internet.

14. Conclusion

Rapid proliferation of AI has impacted many aspects of human life especially in the urban areas. Today, AI is being applied to topics such as urban governance, policy and planning, health care, sustainability, economics and entrepreneurship, etc. However, has several challenges to overcome. As discussed above, efficient telecommunication network is of paramount importance for success of AI in cities. This paper presented a broad overview of these aspects starting with the different topics related to AI in urban life then moving on to the need for advanced telecommunications infrastructure. The focus then shifts to the context of emerging economies moving to AI enabled smart cities. The additional challenges of inadequate telecommunication infrastructure need for AI applications and the socio-economic aspect of providing high quality internet services with affordability meeting high aspirations of the citizens. Study conducted in Kathmandu, the capital city of Nepal, points to the typical challenges emerging economies are likely to face in moving towards AI enabled urban living. The study clearly shows the long bridge needs to be built from the current state of affairs to a modern AI enabled urban life in emerging countries. Future work will try to provide technical solutions to these problems.

**Acknowledgement**

We would to thank Utopia, Kathmandu, particularly, Doma Tsering Tamang and Prabhas Ghimire, for conducting the survey. We also thank Royal Academy of Engineering, UK, for funding this research.